\documentclass[aps,amsmath,amssymb,nofootinbib]{revtex4}

\usepackage{graphics}
\usepackage{bm}
\usepackage{graphicx}
\usepackage{amsmath}
\usepackage{amssymb}
\usepackage{amstext}
\usepackage{amscd}
\usepackage{amsfonts}
\usepackage{appendix}
\usepackage{color}
\usepackage{float}
\DeclareMathAlphabet{\EuFrak}{U}{euf}{m}{n}
\DeclareMathAlphabet{\EuScript}{U}{eus}{m}{n}
\begin{document}

\newcommand{\nd}{\noindent}
\newcommand{\nl}{\newline}
\newcommand{\be}{\begin{equation}}
\newcommand{\ee}{\end{equation}}
\newcommand{\ben}{\begin{eqnarray}}
\newcommand{\een}{\end{eqnarray}}
\newcommand{\nn}{\nonumber \\}
\newcommand{\ii}{\'{\i}}
\newcommand{\pp}{\prime}
\newcommand{\expq}{e_q}
\newcommand{\lnq}{\ln_q}
\newcommand{\quno}{q-1}
\newcommand{\qunoinv}{\frac{1}{q-1}}
\newcommand{\tr}{{\mathrm{Tr}}}
\title{{\bf STATISTICAL COMPLEXITY WITHOUT EXPLICIT REFERENCE TO  UNDERLYING PROBABILITIES
 }}
\author{ F. Pennini$^{1,2}$, A. Plastino$^{3,4}$}

\affiliation{$^{1}$ Departamento de F\'{\i}sica, Universidad
Cat\'olica del Norte, Av.~Angamos~0610, Antofagasta,
Chile\\$^{2}$Departamento de F\'{\i}sica, Facultad de
Ciencias Exactas y Naturales, Universidad Nacional de La Pampa, CONICET, Av. Peru 151, 6300, Santa Rosa, La Pampa,
Argentina\\$^{3}$Instituto de F\'{\i}sica La Plata--CCT-CONICET,
Universidad Nacional de La Plata, C.C.~727, 1900, La Plata,
Argentina\\$^{4}$ SThAR - EPFL, Lausanne, Switzerland
 }
\date{\today}

\begin{abstract}

We show that extremely simple systems of a not too large number of particles  can be simultaneously thermally stable and complex. To such an end, we extend the statistical complexity's notion to simple  configurations of non-interacting particles, without appeal to probabilities, and discuss configurational properties.

\nd Keywords: Disequilibrium without probabilities, Statistical complexity, Disequilibrium.
\end{abstract}

\setcounter{equation}{0}
\maketitle

\section{Introduction}

\nd Examples of collective phenomena which can emerge from real-world
complex systems include traffic congestion, financial market
crashes, wars, and epidemics. They do not involve $10^{24}$ objects or agents but rather thousands, so that canonical ensemble considerations may not be adequate while micro-canonical (MC) ones may apply. MC
dealings do not involve probabilities and will be the focus of our interest here.\vskip 3mm

\nd Now, to be aware of a system's  degree of randomness is not enough for an adequate insight into the extant
  correlation structures. One may try to search for a way to be in a position to discern  the
relations among  a system's components by recourse to a specific quantifier, mimicking the manner in which  entropy $S$ describes disorder.
 Two extreme situations may be encountered:  (a) perfect order or (b)
maximal randomness. In these cases
   strong
correlations do not exist \cite{LMC}. In between,  variegated  degrees of correlation are possible and
we would wish that the above mentioned quantifier would quantify them. We may call it a ``complexity''.  How is one to represent it?  The answer is not easily found.
 Famously,  Seth Lloyd enumerated some 40 manners of defining this ``complexity'',
none of them  optimal.\vskip 3mm

\nd A system may be regarded as complex, obviously, when it does not fit simple patterns, as in the case of either (I) a perfect crystal or (II) the isolated ideal gas. These are good examples of simplicity, or, alternatively, instances of null complexity. In a crystal, the information, or negentropy ($-S$) stored is minimal. A few parameters suffice for a good description. On the contrary, the  ideal gas is completely disordered, with any of its accessible states endowed with the same probability, that exhibits  maximum entropy. Systems (I) and (II) are extreme in the scale of order/information, which implies  that complexity  cannot be cast in  terms of  order or information. In Ref. \cite{LMC} the authors advance  a measure of complexity by employing some kind of distance to the maximum entropy situation, called disequilibrium~(D)~\cite{LRuiz2001}.
  $D$  yields a notion of  hierarchy that would be different from zero if there are privileged states among those accessible ones. $D$ would be maximal for the perfect crystal and vanish for the ideal gas.  For the entropy (S), things are exactly reversed, being minimal for the crystal and maximal for the ideal gas.  Accordingly, L. Ruiz, Mancini, and Calvet~(LMC)~\cite{LMC} advanced, in what constituted a great leap forward,  a statistical complexity measure $C$  of the form

    \be     C=DS\label{primera},\ee
that is an interesting  functional of the probability
distributions that does grasp correlations in the way that entropy
captures randomness~\cite{LMC}.  The quantity $D$ measures (in
probability space) the distance from i) the prevailing probability
distribution  to ii) the uniform probability and it reveals the
amount of structural details~\cite{LMC,cuatro}. For a system of
$N-$particles one has \be
D=\sum_{i=1}^N\,\left(p_i-\frac{1}{N}\right)^2.\label{dlopezR} \ee
Here ${p_1, p_2,\ldots, p_N}$ are the individual normalized
probabilities ($\sum_{i=1}^N\,p_i=1$)~\cite{LMC}. $D$ attains the
maximum value for a fully ordered state and vanishes in the case
of completely disordered states or equiprobable states. Moreover,
LMC's statistical complexity  also it individualize and quantify
the bequeath of Boltzmann's entropy (or information
$S=-\sum_{i=1}^N\,p_i \ln p_i$).

     This proposal received considerable attention and great interest
\cite{LMC,MPR,lmc1,lmc11,lmc2,lmc22,lmc3,cuatro}, being applied in different scenarios for both the canonical and grand canonical ensembles. It is obvious that $C$ vanishes
in the (opposite) simple cases (I)  and (II) above.\vskip 3mm

\nd We will here focus attention, for the first time as far as we know,  on a $C-$scenario devoid of both  i) probabilities and ii) interacting particles. We show that quantifiers like $S$, $D$, and $C$ can be defined and yield new information regarding the behavior of the simplest conceivable system: two level systems of $N$ identical, independent particles. Its simplicity notwithstanding, such systems do exhibit complex behavior and can attain internal stability. It has been forcefully argued that binary decision problems provide the perfect illustration
of complexity \cite{three}, while presenting workers in the field with an exceedingly challenging problem that,as far as we know, no satisfactory mathematics to account for it \cite{three}. Here we consider binary decisions as those of a particle regarding to whether to occupy or not one of the model's two levels. We will obtain intriguing insights.

\nd The paper is organized as follows. In Section~\ref{2level} we introduce relevant concepts related to the two-level model, as well as the motivation of our research. In Section~\ref{defining} we define the quantities $C$ and $D$ mentioned above without using probabilities. This constitutes a crucial issue of endeavor. Section~\ref{sectionfree} is devoted to the study of the free energy $F$, indispensable to establish the existence of regions of thermal stability. Useful relations between $C$ and $D$, together with the specific heat, are obtained in~Section~\ref{relCD}. Possible generalizations are outlined in Section~\ref{generalize}.  Finally, we draw conclusions in Section \ref{conclu}.

\section{Two level system}
\label{2level}
\begin{figure}[H]
\begin{center}
\includegraphics[scale=0.5,angle=0]{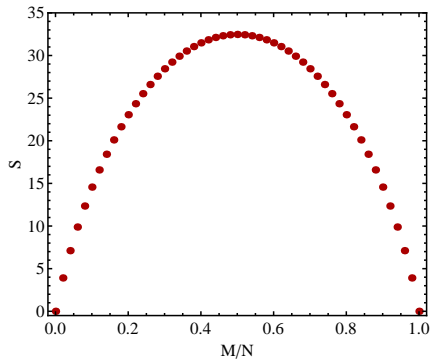}
\vspace{-0.2cm} \caption{ Entropy $S$ as a function of~$M/N$ for $N=50$ in which
$M/N$ runs between 0 and $1$.}\label{figentro}
\end{center}
\end{figure}

\nd We begin by noting that an early complexity-related effort on this theme is due to L. Ruiz \cite{lr2} by considering a laser of two levels of energy and appealing to a normal probability distribution in the canonical ensemble. This valuable work is totally unrelated to the present one, though.\vskip 2mm
\nd Let us consider a system with $N$ indistinguishable particles  with energy $E$ \cite{miranda}. Each particle can be found in two possible states of energy (in suitable units)  $0$ and $1$, respectively, with $M$ particles in the latter level and $N-M$ in the former. In such a scenario, the system's configuration is uniquely characterized by the pair ($N,M$) and has   a state-degeneracy  given by \cite{kelly,kittel,ramsey}

\be
\Omega(M,N)=\frac{N!}{(N-M)!M!},\label{twomicro}
\ee
with, obviously, $M=E$.
Accordingly, following the celebrated equation $S=\ln{\Omega}$ engraved in L. Boltzmann's tomb at Vienna's cemetery,   the entropy obtained from Eq. (\ref{twomicro}) is  (in Boltzmann's constant $k_B$-units)
\be
S= \ln \left(\frac{N!}{(N-M)!M!}\right).\label{entroptwo}
\ee
The general aspect of this entropy is illustrated by Fig.~\ref{figentro} for $N=50$,  not a novel graph that we include here for pedagogical purposes.

\nd  We deal with a collection of configurations (we may speak of micro-canonical configurations) uniquely characterized by the pair ($N,M$). All relevant physical quantities are determined by these pairs. For each of them, i.e., for each possible configuration,  we have fixed values of temperature $T$, disequilibrium $D$, entropy $S$, statistical complexity $C$, and free energy~$F$. The pertinent relationships will be given below in $N,\,M$-terms. This is a particular instance of a much more general one in which the temperature of a
system affects the configurations adopted by that system, and consequently one would expect
to be able to calculate the temperature from configurational information \cite{configurat}.\vskip 3mm

\nd In this work, we are going to compare amongst these configurations and try to discern patterns. Note that, since  for each configuration $T$ is effectively fixed, the free energy concept does make sense. In other words, this paper is a statistical study of the variables $N$ and $M$ over a collection of ($N,M$) configurations.

\subsection{Temperatures}
\begin{figure}[H]
\begin{center}
\includegraphics[scale=0.5,angle=0]{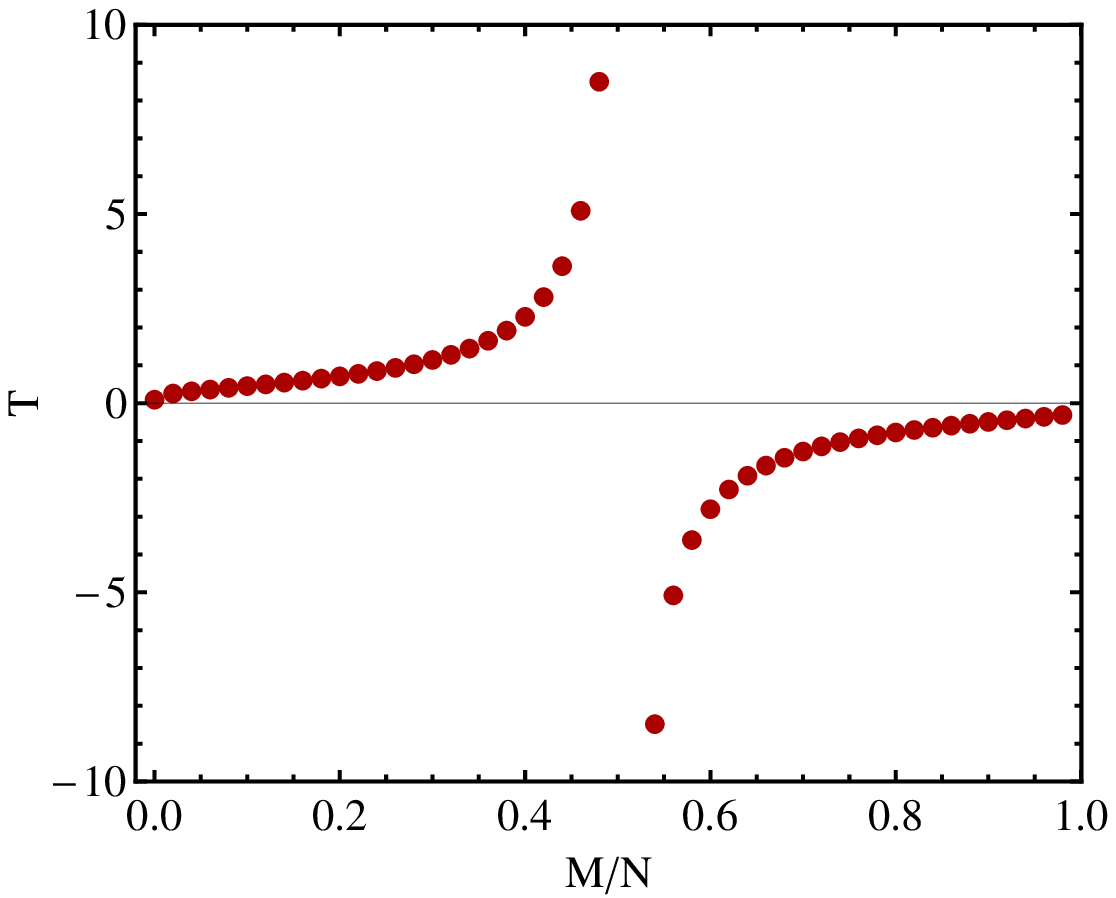}{\small (a)}\,\,\includegraphics[scale=0.5,angle=0]{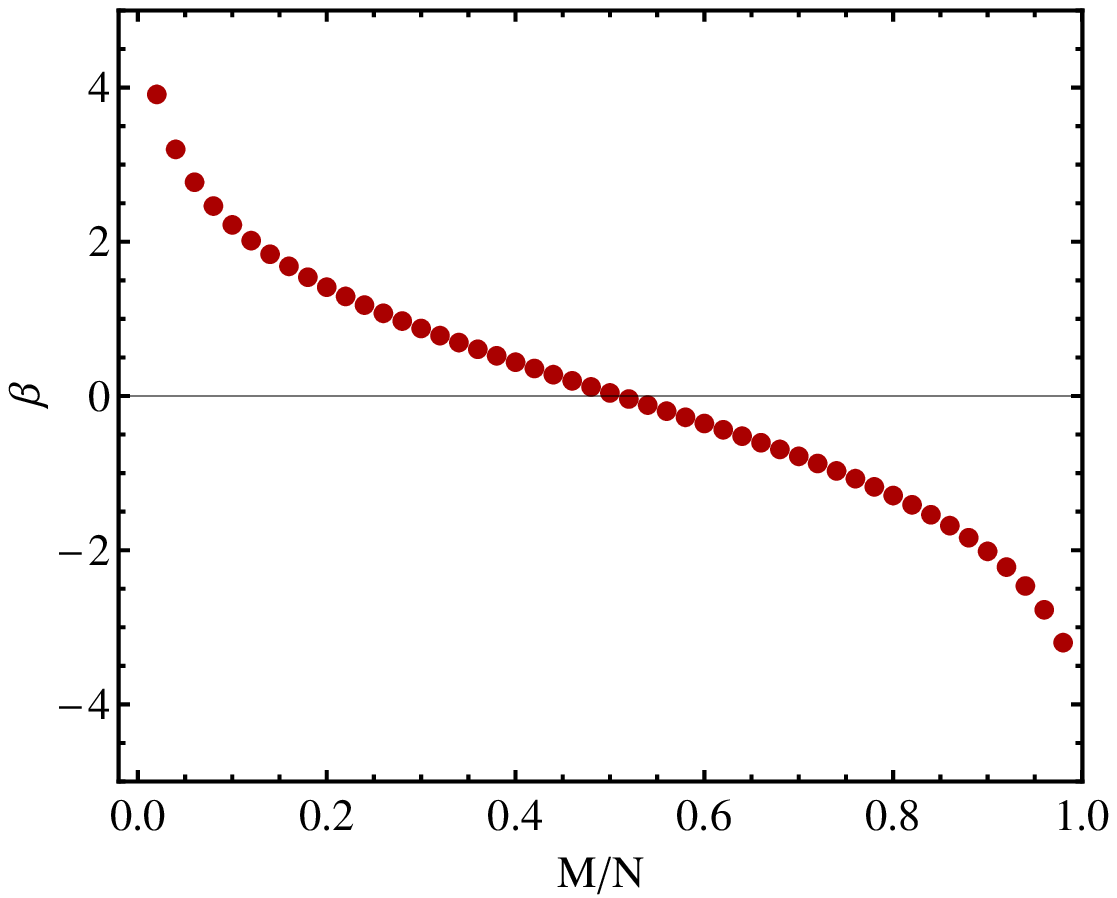}{\small (b)}
\vspace{-0.2cm} \caption{(a): Temperature $T$ versus $M/N$ for $N= 50$. A divergence takes place at $M/N= 0.51$. (b): Inverse temperature $\beta=1/T$ versus $M/N$ for $N=50$.}\label{figtemp}
\end{center}
\end{figure}

The temperature of the system as a function of $M$, $N$ is obtained from the thermodynamical entropy (\ref{entroptwo}). Thus, in this case, one has~\cite{Huang}

\be
\frac{1}{T}=\frac{\partial S(N,M)}{\partial M}.\label{tempmicro}
\ee
Since the energy is a discrete quantity, for not too large systems, the derivatives are replaced by finite differences
and Eq.~(\ref{tempmicro}) becomes~\cite{miranda}

\be
\frac{1}{T(N,M)}= \frac{S(N,M) - S(N,M - 1)}{M - (M - 1)}= S(N,M) - S(N,M - 1).\label{finita}
\ee
Replacing Eq. (\ref{entroptwo}) into Eq. (\ref{finita}), one obtains~\cite{miranda}

\be
\frac{1}{T}=\ln \left(\frac{N}{M}-1+\frac{1}{M}\right),\label{temp}
\ee
entailing that $T$ diverges for $M= (N+1)/2$. An illustration, we depict in Fig.~\ref{figtemp} the temperature  $T$  given by Eq.~(\ref{temp}) for $N=50$. By inverting the above relation we find

\be
M(N,T)=\frac{N+1}{1+e^{1/T}},\label{mt}
\ee
yielding $M(N,T)=E(N,T)$ as a function of $T$.

\subsection{Motivation}
\label{motiv}

\nd Let us reiterate: from Eq. (\ref{temp}), we observe that $T\geq 0$ whenever $N/M-1+1/M> 1$, that is, $M\leq (N+1)/2$. Otherwise, $T< 0$ implies that $M> (N+1)/2$. Moreover, $T$ diverges when $N/M-1+1/M=1$ implying that $M=(N+1)/2$.
It is well known that negative temperatures always arise if there is an upper bound to the system's energy. States with negative temperature are actually hotter than states with positive temperature \cite{kelly,kittel,ramsey}. The $T-$divergence is the origin of the interesting physics to be described below. Actually, Fig.~\ref{figtemp} constitutes our motivation for investigating the statistical complexity in this context.
 It is clear that the above mentioned divergence may justify hopes of finding in it the source of complex behavior. In particular, note that, trivially, $\beta=1/T$ behaves as an order parameter \cite{glass} in the following sense:
 let us call ``symmetric'' (occupationally symmetric) that state with roughly the same number of particles in each of the two levels. $\beta$ is zero  in the high temperature, or symmetric, state, but at low temperatures, when this symmetry is
  ``broken'', it takes on a nonzero value \cite{glass}. The novel point here is that we will show below that
     LMC's  disequilibrium $D$ is also an order parameter in this peculiar sense. \vskip 3mm

\nd On a different vein, and as stated in the Introduction, in this paper we deal, in essence, with a collection of micro-canonical configurations. Why to look at complexity in such a setting?  Because it seems unlikely that theories which need to assume
very large numbers of objects (so that appeal to the canonical ensemble makes sense) may represent everyday
complex systems, where the numbers involved are typically less
than a thousand, or even a hundred \cite{three}. For instance, in a financial
market the number $N_P$ of people who actually have enough economic
clout to ``move'' the market is relatively small,  and it is this $N_P$  which should feature in any realistic model of the market  \cite{three}. Thus, micro-canonical modeling of complexity is indeed reasonable and does not involve probabilities.

\section{Defining   disequilibrium and statistical complexity without an underlying probability distribution}
\label{defining}

\nd This is the crucial issue. Once we have an adequate
``probability-less'' $D-$version everything follows smoothly.
Given~$N$, we have a collection of $M+1$ (different)  possible
energy-configurations. Elementary combinatorial arguments show
that the configuration of maximum entropy (ME) is that of the pair
$(N,N/2)$ if $N$ is even. For $N-$odd, ME is attained, with equal
values,  at $M= N/2 \pm 1/2$.

\begin{figure}[H]
\begin{center}
\includegraphics[scale=0.5,angle=0]{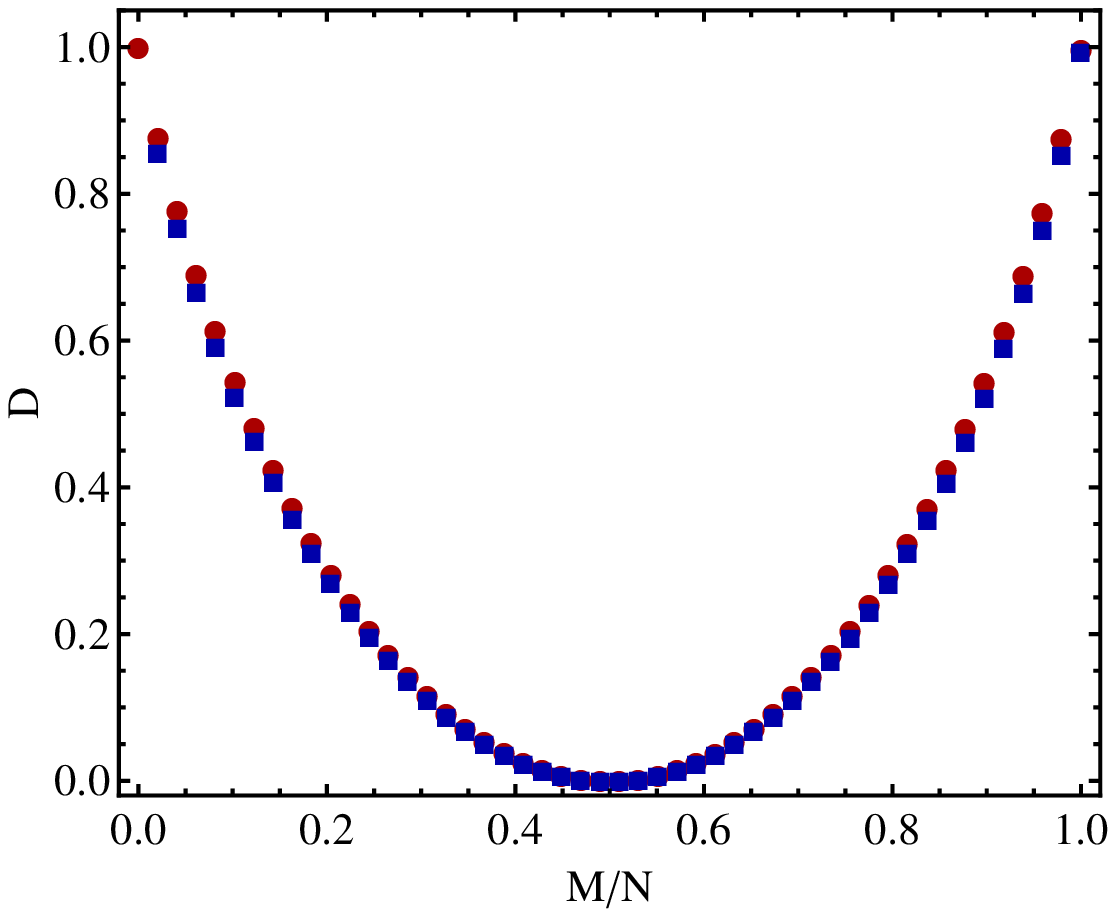}{\small (a)}\,\,\includegraphics[scale=0.5,angle=0]{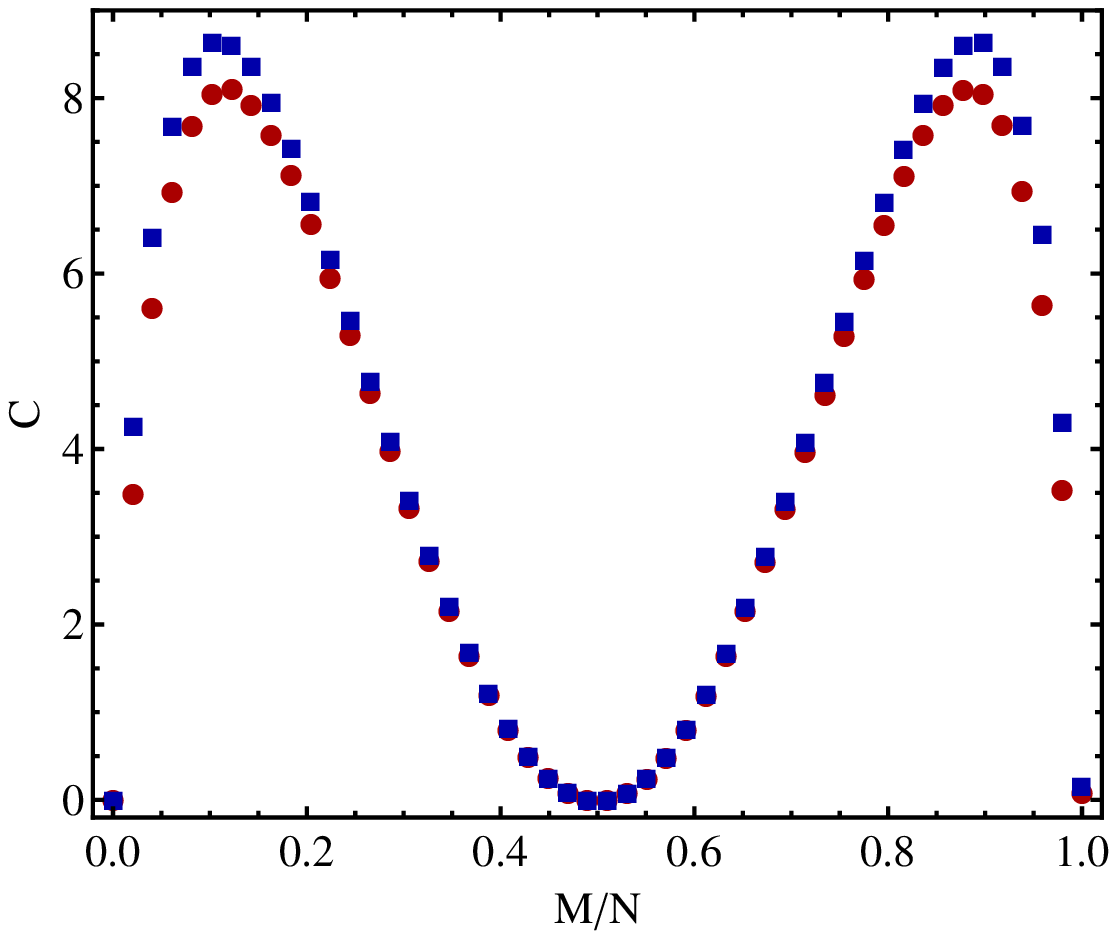}{\small (b)}
\vspace{-0.2cm} \caption{(a): Disequilibrium $D$  as a function of~$M/N$ for the  surrogate probabilistic distribution (blue square) of Eq. (\ref{bep}), compared to this paper's definition (red circle) for $N=50$.  (b): Same as Fig. (a) but for the  statistical complexity $C$. In this case, the maxima are in $M/N=6/50$ and $M/N=44/50$.}\label{comparaD}
\end{center}
\end{figure}

\nd LMC defined $D$ as a distance in probability space: that to the uniform distribution (UD). We work here in a scenario devoid of probabilities. What to do?  Noting that the UD is the maximum entropy ME-distribution, the judicious $D-$choice is then the following: the disequilibrium  $D$ of the configuration $(N,M)$ is to be properly defined as a kind of ``distance'' to  the ME configuration (in configuration space). We choose, given the ME value $ S_{max}$,

\be
D(N,M)=1-\frac{S(N,M)}{S_{max}(N,M)},\label{dis1}
\ee
where $S_{max}(N,M)=S(N,N/2)=\ln(N!/(N/2)!^2)$. It vanishes for $S=S_{max}$ and is maximal for $S=0$, as one should expect. The corresponding analytic expression is

\be
D(N,M)=\frac{\ln\left(\frac{(N-M)!M!}{(N/2)!^2}\right)}{\ln\left(\frac{N!}{(N/2)!^2}\right)}.
\ee
According to (\ref{primera}), we have now a probability-less LMC complexity given by
\be
C(N,M)=D(N,M) S(N,M).\label{compleLMC}
\ee
The adequacy of our $D-$definition could be assessed by comparing it to an orthodox LMC one that uses a suitable probability distribution. This is the goal of  Fig.~\ref{comparaD}, that  compares, for $N=50$, our $D$ versus $M/N$ with  the orthodox LMC disequilibrium for a surrogate Boltzmann-like exponential distribution (BEP), with good agreement. This surrogate BEP is, for a given $N$, of the form

\be  P_B= \exp{(-E_M/T_M)}/Z;\,\,\,Z=\sum_M\,\exp{(-E_M/T_M)},     \label{bep}\ee
where $T_M$ is the temperature of the $(N,M)$ configuration. Of course, a ``true" Boltzmann distribution has a common $T$ for all $M$.  We use $P_B$ so as to construct an orthodox LMC-$D$ and, by comparison, validate the reasonability of our probability-less $D$-definition. Beyond this role, we suggest that $P_B$ might arouse some interest by itself, though. \vskip 1mm
\vskip 2mm \nd We remind the reader that the true Boltzmann
distribution exhibits the appearance

  \be  P_B(M)= \exp{(-E_M/T)}/Z;\,\,\,Z=\sum_M\,\exp{(-E_M/T)},
\label{bep1}\ee with an entropy $S_B$

\be  S= -\sum_M P_B(M) \ln{P_B(M)}, \label{bep2}\ee and a
disequilibrium $D$ equal to (see (\ref{dlopezR}))

\be  D= \sum_M [P_B(M)- (1/M)]^2. \label{bep3}\ee The complexity
$C$ is then usually evaluated using Eq. (\ref{primera}).
 Comparing (\ref{bep}) with (\ref{bep1}) we see
that our surrogate distribution is just a  Boltzmann  probability
distribution with an $M-$dependent temperature. \vskip 2mm

\nd Also, note that our $D$ is zero at very high temperatures (the symmetric state defined in Subsection~\ref{motiv}), and grows as the temperature descends, behaving thus like an order parameter \cite{glass} and becoming then a thermodynamic variable~\cite{glass}.

\nd Also, we see that $C$ vanishes for maximum entropy and for
zero entropy, as it should. It also exhibits two peaks.
  As
stated above  in Subsection~\ref{motiv}, $D$ plays the role of an
order parameter in the sense therein explained. Accordingly, LMC's
statistical complexity becomes the product of two thermodynamic
variables, something that, we believe, has not been remarked
before. Moreover, we encounter complexity peaks for very small
values of $N$. Indeed, this happens already for $N=3$, which
brings to mind the title of \cite{three}: {\it Two's Company,
Three is Complexity}. In order to ascertain the $C-$peaks'
significance we turn next to the free energy $F$. We will see that
only the first maximum is physically relevant.

\section{The free energy $F$}
\label{sectionfree}
\nd This is an essential quantity which is defined, for $E=M$ (because the lowest of our two levels has zero energy), as

\be
F(N,M)=M-T S(N,M).
\ee
 The important  point here is that $\mathrm{d}^2 F/\mathrm{d}T^2$ should be $<0$ for thermal stability~\cite{callen} (Appendix G,
Eqs.~G2, G7, G8, and G9).  Explicitly, one finds in this celebrated text-book that

\be \frac{\mathrm{d}^2 F}{\mathrm{d}T^2}= -\frac{C_h}{T}, \label{shf}\ee
with $C_h$ standing for the specific heat, a positive quantity (except for gravitational systems \cite{lynden}). Thus, in our regions of positive temperature,
 $\mathrm{d}^2 F/\mathrm{d}T^2$ is negative.
 Additionally,  we observe that for the two-level system,  $\mathcal{C}_h$ is given by the discrete relation~\cite{miranda}

\be
\mathcal{C}_h(N,M)=\frac{1}{T(N,M)-T(N,M-1)}.
\ee
If we replace this into Eq. (\ref{shf}) we arrive at

\be
\frac{\mathrm{d}^2 F}{\mathrm{d}T^2}=-\frac{1}{T(N,M)[T(N,M)-T(N,M-1)]}.\label{d2t}
\ee
 \vskip 2mm

\begin{figure}[H]
\begin{center}
\includegraphics[scale=0.5,angle=0]{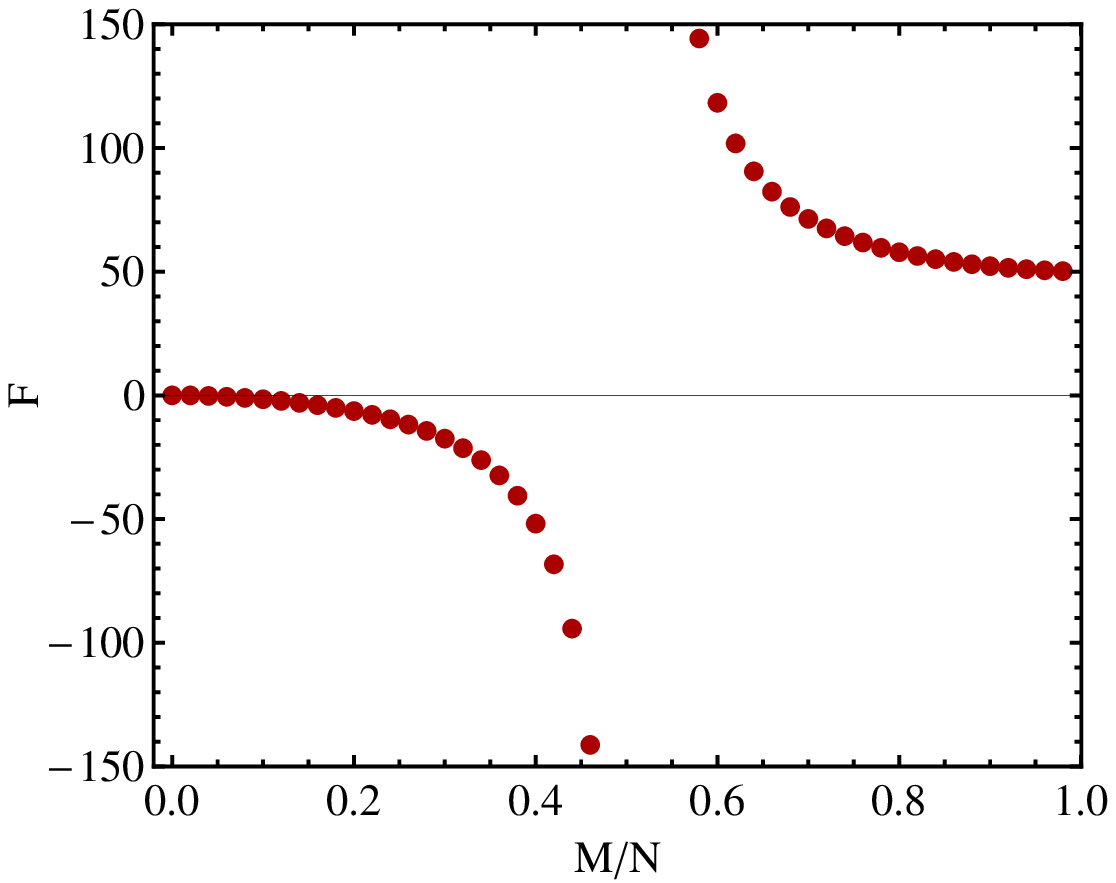}{\small (a)}\,\,\includegraphics[scale=0.5,angle=0]{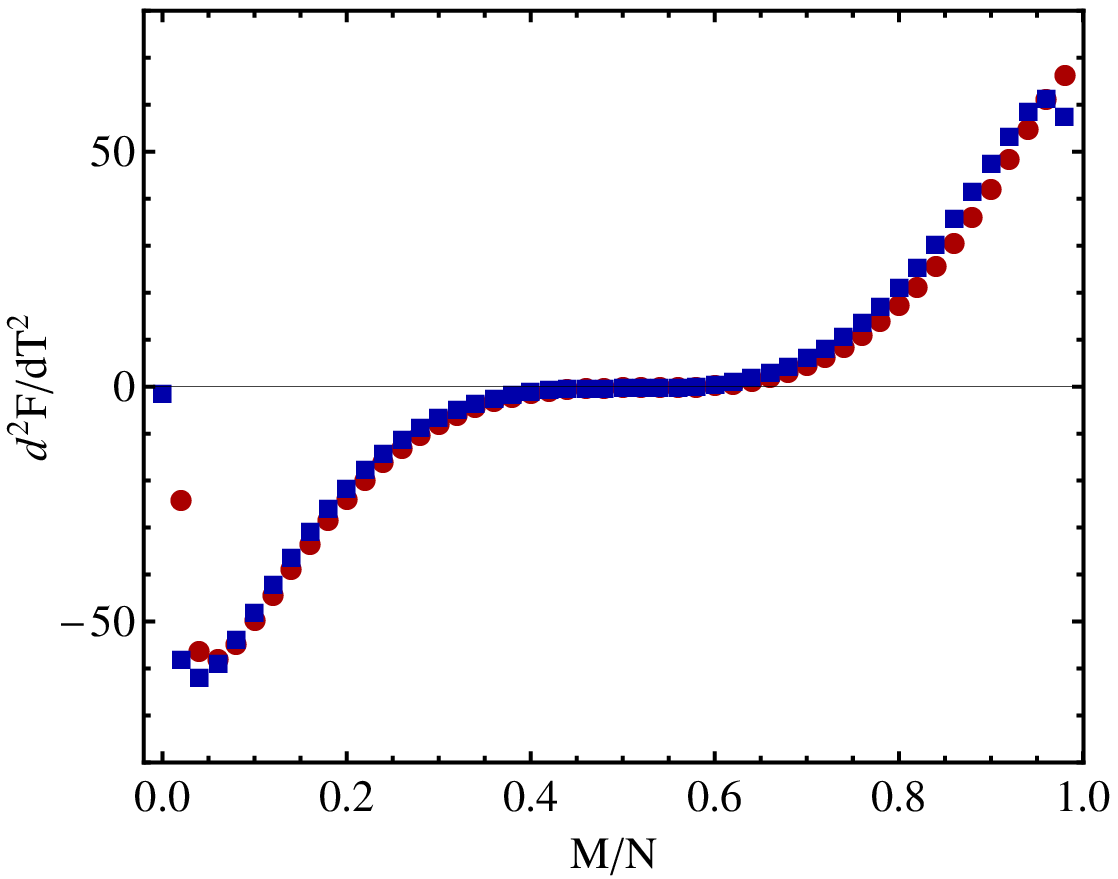}{\small (b)}
\vspace{-0.2cm} \caption{(a): Free energy $F$ as a function of~$M/N$ for $N=50$. Of course, $F$ diverges at the same place at which $T$ does.  (b):  second derivative of the free energy $F$ with respect to temperature $T$ as a function of~ $M/N$ for $N= 50$. The blue line corresponds to $\mathrm{d}^2F/\mathrm{d}T^2$ for the two-level system using the surrogate distribution (\ref{bep}) in constructing $F$.}\label{free_condition}
\end{center}
\end{figure}

\nd  In Fig. \ref{free_condition} we plot 1) $F$ vs. $M/N$ (left) and 2) $\mathrm{d}^2 F/\mathrm{d}T^2$ (given by  Eq. (\ref{d2t})) vs. $M/N$ (right). This second derivative is $<0$ only in the region of the first  of the two complexity maxima, which becomes then the only physically relevant one. Our recurrent divergence reappears for $F$ at the proper  $M/N$ place. A $\mathrm{d}^2 F/\mathrm{d}T^2$-minimum is attained in the vicinity of $M/N=0.1$, where one finds the first complexity maximum. One might conjecture that, if $F<0$, the system could be regarded as ``bounded'', since one would need
 to provide energy so as to ``break it up'' \cite{desloge}. On such a vein, one could guess that this should spontaneously happen  for $F>0$. {\it The maximum complexity and the maximum stability take place at roughly the same $M/N$-value. The most complex configuration is the most stable one. It is attained at $T>0$ and $F<0$.}

\begin{figure}[H]
\begin{center}
\includegraphics[scale=0.5,angle=0]{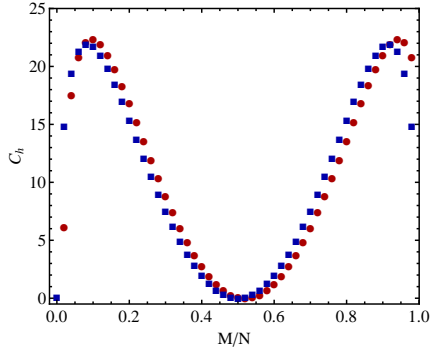}
\vspace{-0.2cm} \caption{Our version of the specific heat $\mathcal{C}_h$  as a function of $M/N$ for $N=50$ ( red). Blue squares: LMC  $\mathcal{C}_h$, using the surrogate distribution (\ref{bep}).}\label{sheat}
\end{center}
\end{figure}

\begin{figure}[H]
\begin{center}
\includegraphics[scale=0.5,angle=0]{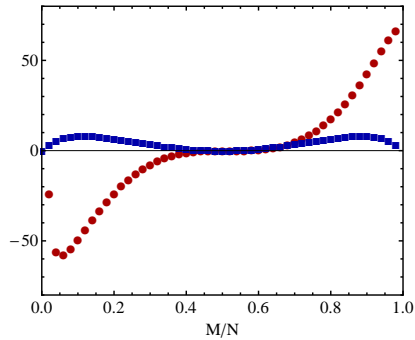}
\vspace{-0.2cm} \caption{$C$ (blue square) and $\mathrm{d}^2F/\mathrm{d}T^2$ (red circle) versus $M/N$ for $N=50$. }\label{dos}
\end{center}
\end{figure}

\nd The specific heat $\mathcal{C}_h$ vs. $M/N$,  is plotted in Fig. \ref{sheat}  together with its version according to the surrogate probability distribution of Eq.~(\ref{bep}). Nest, we depict $C$ and $\mathrm{d}^2F/\mathrm{d}T^2$ versus $M/N$ in Fig.~\ref{dos}, that  exhibits the notable fact that maximum statistical complexity obtains at the same $M/N$-location at which one has maximal thermal stability.

\begin{figure}[H]
\begin{center}
\includegraphics[scale=0.5,angle=0]{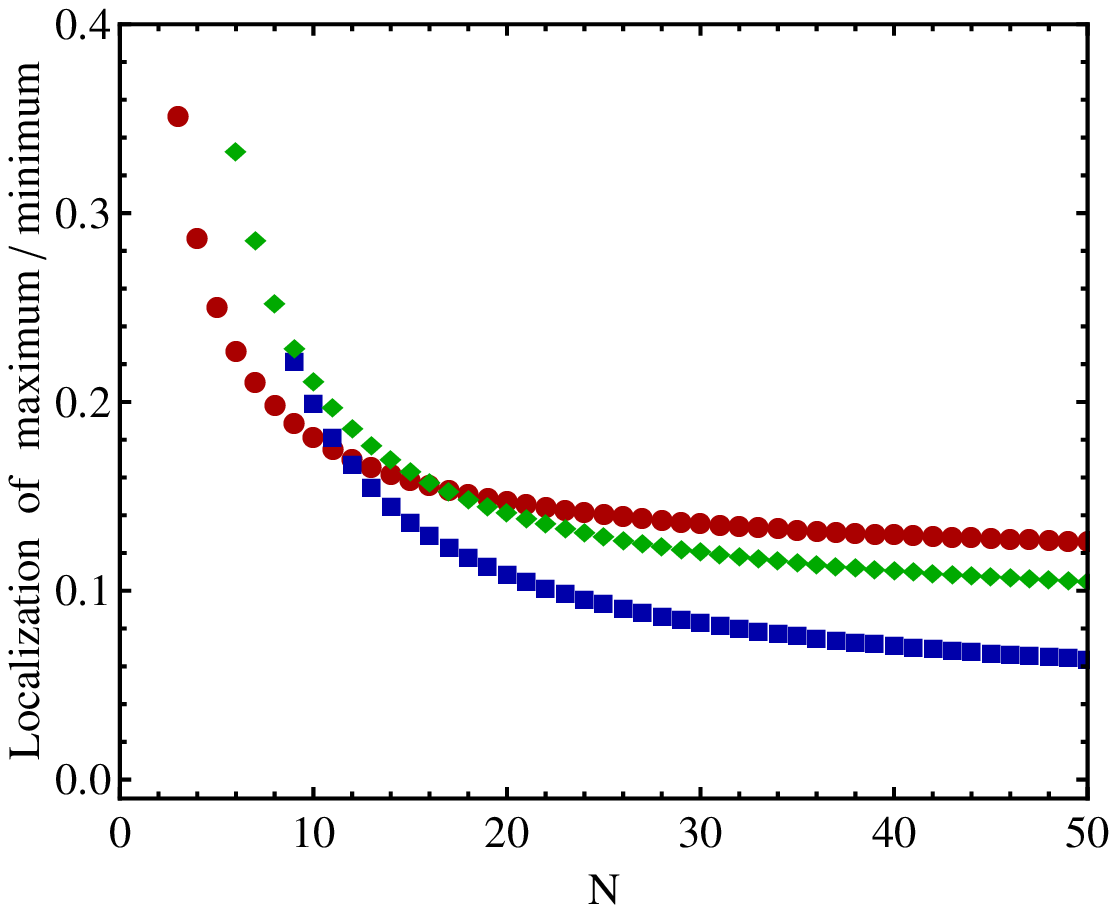}{\small (a)}\,\,\includegraphics[scale=0.5,angle=0]{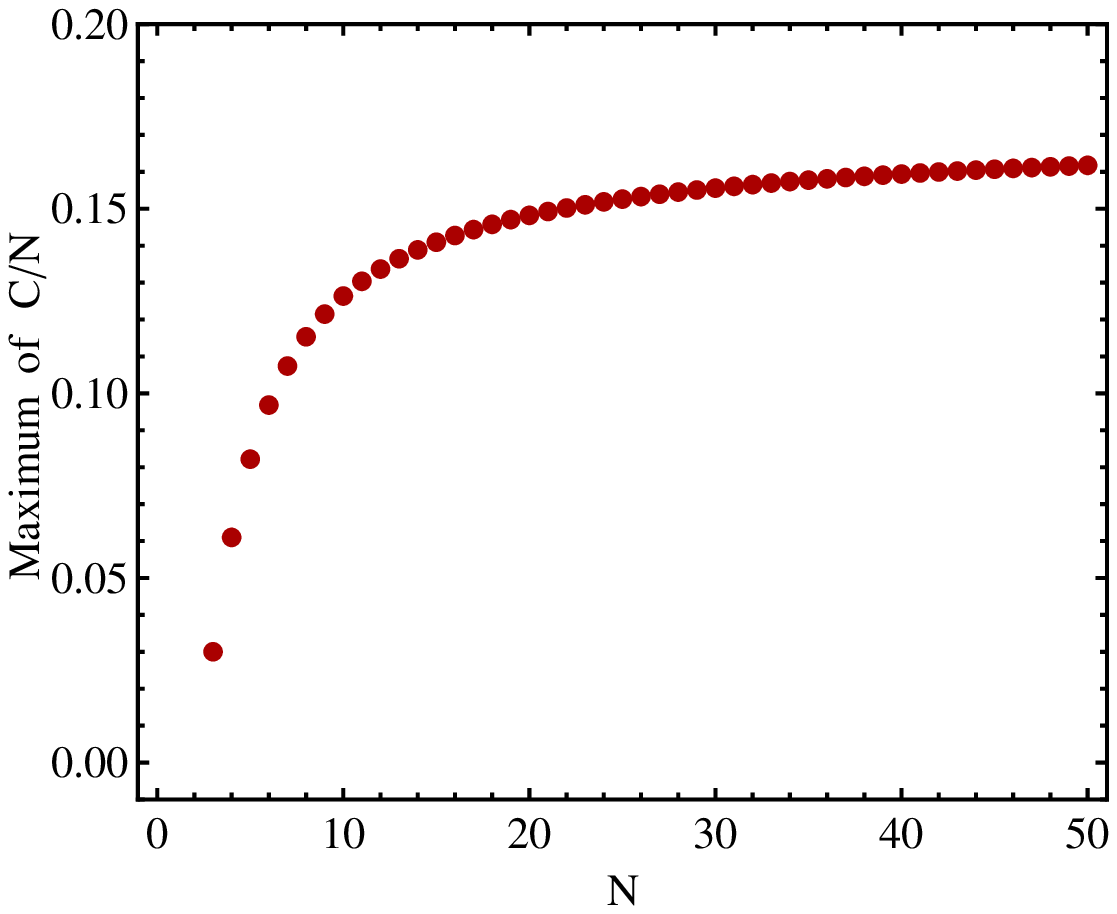}{\small (b)}
\vspace{-0.2cm} \caption{(a): Plot of $M/N-$location of the maximum value of $C$ (red circle), minimum of  $\mathrm{d}^2 F/\mathrm{d}T^2$ (blue square), and maximum of  $\mathcal{C}_h$ (green diamond) as a function of $N$ that runs between 3 and 50. (b): Maximum of $C/N$ versus $N$ between 3 and~50. We see that the maximal complexity per particle is of about $0.17$ for most of the $N-$range.}\label{maxi}
\end{center}
\end{figure}

\nd In addition, we also add here  in Fig. \ref{maxi} a plot depicting the dependence on $N$ of the $M/N$-{\it location} of three important quantities: 1) $C$, 2) $\mathcal{C}_h$, and 3) $\mathrm{d}^2 F/\mathrm{d}T^2$. The last two refer to the most stable situation, while the first does so for the complexity-maximum. The three locations are close neighbors.  Also, maximum complexity and stability are attained at the same location for $N=3$, which brings to mind the book-title in Ref. \cite{three}.

\nd In order to calculate the maximum of $C$ for  $N$  between 2 to 200 particles, we look at  the discrete derivative of $C$ respect to $M$ \cite{miranda}, i.e.,
\be
\frac{\mathrm{d}C(N,M)}{\mathrm{d}M}=C(N,M)-C(N,M-1).
\ee
The extremes are attained when $\mathrm{d}C/\mathrm{d}M=0$. Therefore, employing definitions (\ref{finita}), (\ref{dis1}), and (\ref{compleLMC}), we obtain

\be
S_{max}(N,M)=T(N,M)\left[S^2(N,M)-S^2(N,M-1)\right].
\ee
The above equation is solved numerically to find that $M$ for which $C/N$ is maximum. The results are depicted in Fig.~\ref{maxi}.\vskip 3mm

\nd  The asymptotic behavior is found as follows. First, for $N\gg 200$, we calculate, for fixed $N$,
the first derivative of $C$ with respect to  $M$. One has

\be
\frac{\mathrm{d}C}{\mathrm{d}M}=\left(1-\frac{2 S}{S_{max}}\right) \frac{\mathrm{d}S}{\mathrm{d}M}.
\ee
We see that $\mathrm{d}C/\mathrm{d}M$ is zero provided that $1-2 S/S_{max}=0$ plus $\mathrm{d}S/\mathrm{d}M\neq 0$. Such conditions lead to  $S=S_{max}/2$, so that, replacing this into Eq. (\ref{compleLMC}), and using Stirling's
approximation ($\ln N!\approx N \ln N-N$), we  immediately find

\be
\frac{C_{max}}{N}\approx \frac{\ln 2}{4}=0.1732.
\ee
For $N>200$, $C$ grows linearly with $N$. One might wish to see this result as an extremely simple  instance of  Anderson's {\it  more is different} apothegm \cite{anderson}.

\section{Relations amongst $T$, $C$, $D$, and the specific heat}
\label{relCD}

\begin{figure}[H]
\begin{center}
\includegraphics[scale=0.5,angle=0]{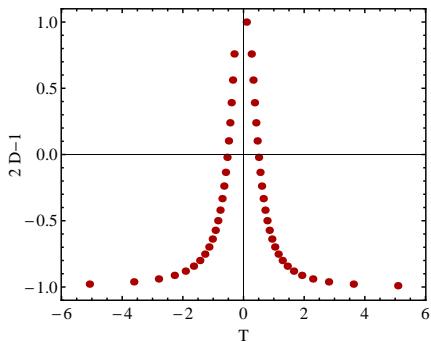}
\vspace{-0.2cm} \caption{$2D-1$ as a function of~temperature $T$ for $N=50$. }\label{disT}
\end{center}
\end{figure}

\nd We begin this section making use of Eq. (\ref{dis1}), from which we obtain  the entropy  as

\be
S=S_{max}\,(1-D).\label{r1}
\ee
Thus, using the relation (\ref{r1}), the specific heat adopts the appearance

\be
\mathcal{C}_h=T \frac{\mathrm{d}S}{\mathrm{d}T}=-S_{max} T\,\frac{\mathrm{d}D}{\mathrm{d}T}.\label{r2}
\ee
Taking into account that the statistical complexity is $C=D S$,  we also get

\be
\frac{\mathrm{d}C}{\mathrm{d}T}=S_{max}(1-2D)\,\frac{\mathrm{d}D}{\mathrm{d}T}.\label{r3}
\ee
Considering Eqs. (\ref{r2}) and (\ref{r3}), we find   a crucial relation  between $\mathcal{C}_h$, $D$, and $C$,  namely,

\be
\mathcal{C}_h=\left(\frac{T}{2 D-1}\right)\frac{\mathrm{d}C}{\mathrm{d}T}.\label{r4}
\ee
Stability of a thermodynamic systems requires $\mathcal{C}_h\geq 0$ so that Eq. (\ref{shf}) tells us that  stability is attained when $\mathrm{d}^2F/\mathrm{d}T^2\leq 0$ at $T\geq 0$ --see discussion in subsection~\ref{sectionfree}. To verify the stability criteria we show, in Fig. \ref{disT}, the behavior of the quantity $2 D-1$ as a function of temperature $T$. There, we see that
  $ 2D-1\geq -1$ for all~$T$. Thus, in order to guarantee stability, two  possibilities arise we are: $2D-1> 0\,\wedge$  $\mathrm{d}C/\mathrm{d}T\geq 0$, and
 $2D-1<0\,\wedge$  $\mathrm{d}C/\mathrm{d}T< 0$. Since we are here interested in the region  $T>0$, this implies that $0\leq M/N\leq0.5$.

\begin{figure}[H]
\begin{center}
\includegraphics[scale=0.5,angle=0]{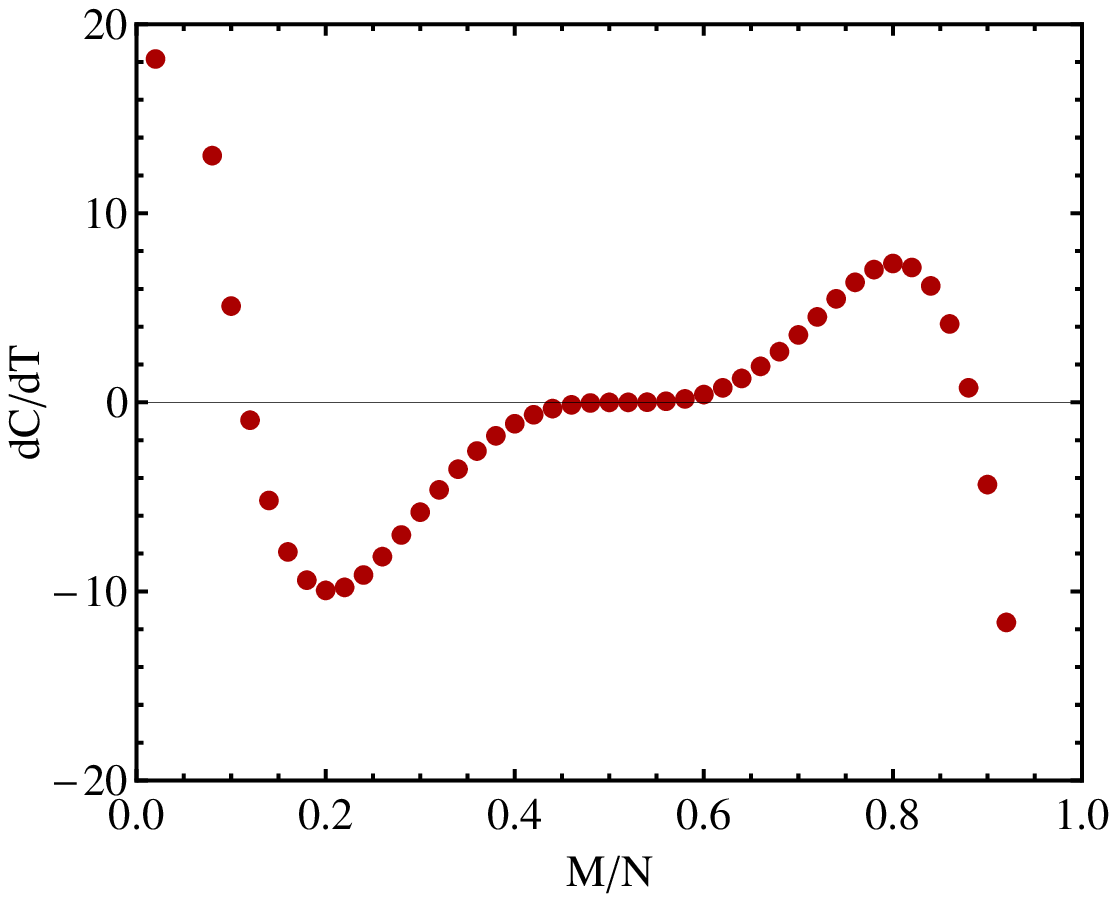}{\small (a)}\,\,\includegraphics[scale=0.5,angle=0]{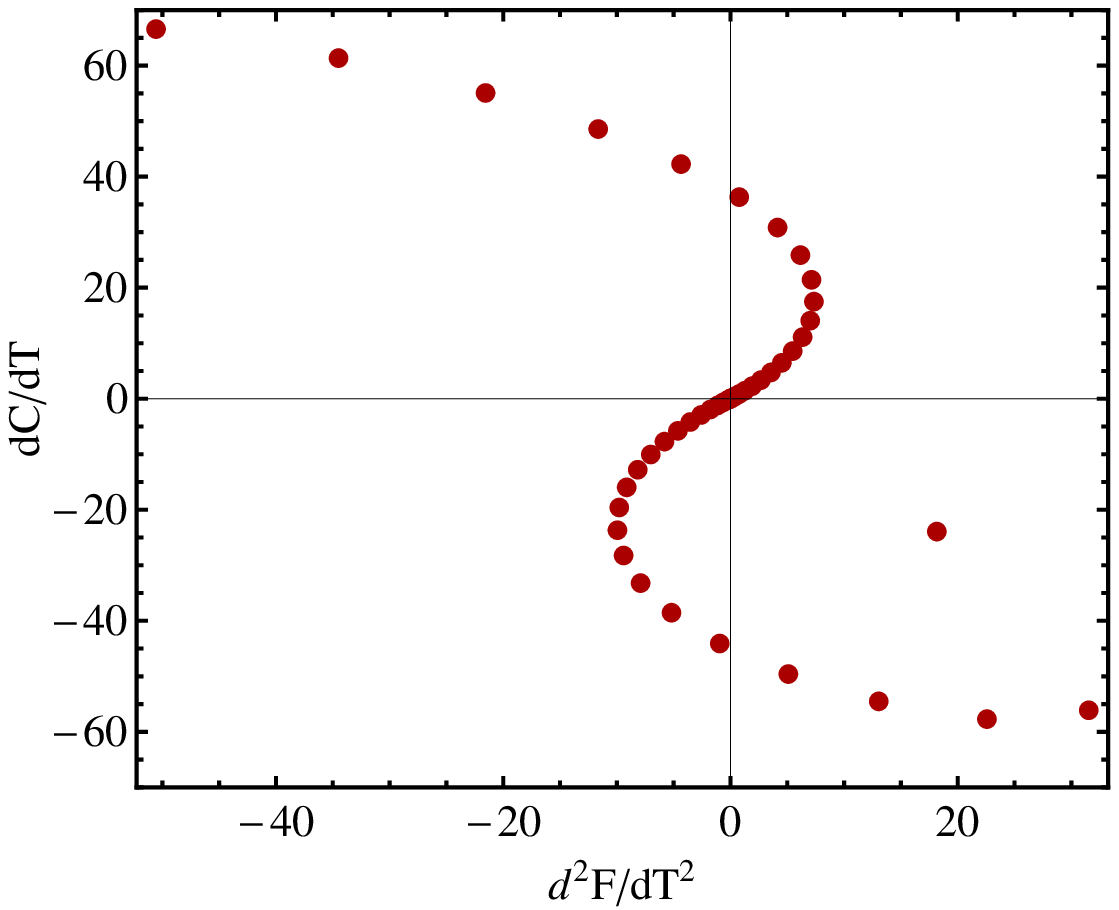}{\small (b)}
\vspace{-0.2cm} \caption{(a): $\mathrm{d}C/\mathrm{d}T$ versus $M/N$ for $N=50$. (b): $\mathrm{d}C/\mathrm{d}T$ versus $\mathrm{d}^2F/\mathrm{d}T^2$ for $N=50$. We note the stability region covers the second and third quadrant.}\label{figucriter}
\end{center}
\end{figure}

 \nd  In addition, according to  Eq. (\ref{r2}), stability is also attained when $\mathrm{d}D/\mathrm{d}T\leq 0$ implying that, by Eq.~(\ref{r1}), $\mathrm{d}S/\mathrm{d}T\geq 0$, as it should. Summing up, the stability criterion for this simple system, which we illustrate in Fig. \ref{figucriter}, is

\be
\mathcal{C}_h\geq 0 \Leftrightarrow \mathrm{d}^2F/\mathrm{d}T^2\leq 0\Leftrightarrow \mathrm{d}C/\mathrm{d}T\leq 0\,\,\, \textrm{for}\,\,\,0\leq M/N\leq0.5.
\ee
It is clear that the complexity should diminish as the temperature grows, as expressed by the relation
 $ \mathrm{d}C/\mathrm{d}T\leq 0$. The coherence of our overall picture can be appreciated once again. This common-sense observation becomes here elevated to the status of stability criterion.

\section{Possible generalizations beyond two-level systems}
\label{generalize}
\nd For tackling more realistic scenarios, the formalism presented above needs generalization. Now, in order to generalize the LMC measure proposed here to more realistic scenarios one must require that they exhibit quantifiable features. This is indeed  possible in the cases of collective phenomena which  emerge from real-world complex systems like traffic congestion, financial market
crashes, wars, cancer,  epidemics, etc. (see Ref.~\cite{three}). If one has access to a measure $I$ that quantifies our ignorance with respect to some relevant aspects of the phenomena under scrutiny~\cite{jaynes}, then one can also assess its maximum possible possible value $I_{max}$ and construct a suitable LMC-like statistical complexity of the general form:

\be \label{general} C_I= (1-I/I_{max}) I,\ee
a suitable generalization of (\ref{compleLMC}). The foremost ignorance measure is Shannon's one $H$
\cite{jaynes}, so that

\be \label{shannon} C_{Shannon}= (1-H/H_{max}) H,\ee that does involve probabilities. However, these probabilities need not be the ones associated to Gibbs-ensembles. An example are the surrogate probabilities~(\ref{bep}) above.  This might be feasible for some of the complex scenarios  envisaged in Ref.~\cite{three} and could then generate future research.

\section{Conclusions}
\label{conclu}
\nd Most collective phenomena emerging from actual complex systems, like traffic jams,
financial market moves, different types of conflicts etc.,  do not involve $10^{24}$ objects or agents but rather thousands, so that canonical ensemble considerations should better be replaced by   micro-canonical ones, that do not involve probabilities. The successful complexity quantifier called statistical complexity assumes an underlying probability distribution, so that for using it in the above indicated scenarios it has to be adapted to a framework without probabilities. This is what we have done here for one of the simplest conceivable physical system: the two levels model.\vskip 1mm

\nd Form another angle, note that binary decision problems represent a common scenario
that yields a paramount example
of real-world complexity \cite{three}, providing  researchers with
 a  challenging problem
which has,  as far as we know, no exact mathematical theory to accompany it \cite{three}. In this work a modest first step towards it has been taken, by expressing a binary decision in terms of occupying (or not)  one of our two levels and $N$ particles, $M$ of which occupy the highest-lying one. \vskip 2mm
\nd We studied the set of the concomitant two-level configurations  of $N-$particles and constructed statistical quantifiers for it like disequilibrium $D$ and statistical complexity $C$, without appeal to the notion of probability.
\vskip 1mm

\nd Our two-level configurations have fixed $E$ and  $N$.
Our focus was the collective of the $N,\,M$-configurations, each with different but fixed energy, particle-number, entropy, and temperature. We have described the properties of the collective, in particular, the values for $D$ and $C$.
\vskip 1mm
\nd We have shown that $D$ is an order parameter and thus a thermodynamic variable. Accordingly, LMC's statistical complexity became the product of two thermodynamic variables, something that, we believe, had not been remarked before.

\vskip 1mm \nd Since $-d^2 F/dT^2= C_h/T$, we were able to show
that the $M/N-$location of the complexity maxima coincides with
that of the $C_h$ ones, we conclude that the states of maximum
complexity are the most stable ones, a very nice feature.  We have
seen that, for not too large $N-$values, configurations exist that
are simultaneously complex and stable, which is remarkable given
the system's extreme simplicity. \vskip 1mm \nd If one deals with
a set of $N$ micro-canonical configurations labelled by an integer
$M$, as here, each of them endowed with its own temperature $T_M$,
one can successfully treat the set withe the surrogate probability
distribution $P_B$ of Eq.~(\ref{bep}).

\vskip 1mm \nd We hope that our present
close look to the inner workings of  complexity  in
a simple environment will contribute to the elucidation of this notion and may stimulate other researchers to further delve on related issues.

\end{document}